\documentstyle[12pt]{article}
\newcommand{\nf}{{{\mathrm n}}}
\newcommand{\pf}{{{\mathbf p}}}
\newcommand{\Bf}{{{\mathbf B}}}
\newcommand{\Ef}{{{\mathbf E}}}

\newcommand{\e}{{\mathbf{e}}}

\newcommand{\rf}{\mathbf{r}}

\newcommand{\ce}{{\cal{E}}}

\begin{document}

\title{ From light-speed state to the rest: new representation for energy-momentum }

\author{
Robert M. Yamaleev\\
Facultad de Estudios Superiores,\\
Universidad Nacional Autonoma de Mexico\\
Cuautitl\'an Izcalli, Campo 1, C.P.54740, M\'exico.\\
Joint Institute for Nuclear Research, Dubna,Russia\\
 Email:iamaleev@servidor.unam.mx }
 \maketitle
%
%
\begin{abstract}

New representation for energy-momentum of relativistic particle near zero-mass point is found.  Formulae of the
energy-momentum are expressed via counterpart of the rapidity which provides a regularity of the representation near
zero-mass and speed of light points. The representation implies an additional characteristic parameter counterpart to
the inertial mass and admits an extension of the energy-mass equivalence principle to the class of massless particles.
The nature of mass of the "left" neutrino and violation of the parity are explained from the first principles.
Hypothesis on quantization of the energy-momentum and the velocity near the light velocity is suggested. The group of
transformations using the counter-rapidity as a parameter of transformation is constructed. Geometric interpretation of
the counter-rapidity as a distance in hyperbolic plane is given.

\end{abstract}


\section{ Introduction}

The areas near the points of zero-mass and near the speed of light are for relativistic physics the areas of a special
interest because masses of the elementary particles are either quite small or equal to zero.
The state of a relativistic particle with zero inertial mass and with velocity equal to speed of light is a
peculiar point of the classical relativistic dynamics. In fact, at this point the energy and momentum of a massive
particle become infinity, which it follows from formulae
$$
cP_0=\frac{mc^2}{\sqrt{1-v^2/c^2}},~~P=\frac{mv}{\sqrt{1-v^2/c^2}}.
$$
It is seen, the point $v=c$ is a singular point. At this point the proper mass of any particle has to be equal to zero.
Thus, the singularity can be resolved by tending the proper mass to zero simultaneously tending the velocity to the
speed of light. It is expected then, at this common limit one will obtain some finite value of the energy (momentum)
corresponding to the energy (momentum) of some massless particle. However, the classical relativistic dynamics
does not provide one with such a rule.

The relativistic kinematics deals only with the velocities of the particles and the observers. The inertial mass is the
very parameter of an inertial system which does not admit to attain the speed of light. For the "inertial observes"
this point is prohibited by convention. The state of the light-speed is a fixed point. Within the framework of the
classical theory the massive particle continuously accelerating may only approach to the light-speed state. As soon as a particle
falls into the light-speed state it cannot escape this state because of the addition formula for the velocity
$$
v'=\frac{v+V}{1+vV/c^2}.
$$
According to this formula, if initial state is equal to speed of light, $v=c$, then the velocity at the final state also is given by this
value: $v'=c$. Here, it is important to take into account that this state is allowed only for particles with zero
inertial mass. This means, in order to escape this state one has to be able to add some quantity of the inertial mass
into the "imaginary engine" confined inside the light-speed state. In the other words,  in order to avoid the state of the light speed
one has to be able to vary the proper mass of the particle.

The Lorentz group of transformations as a parameter of the Lorentz-boost uses an hyperbolic angle, the {\it
rapidity}. Most physicists maintained that the rapidity is a merely formal quantity introduced a purely mathematical
device. The others meanwhile, insisted that the limitation $v<c$ and the non-additivity of the velocity parameter $v$
required the rapidity to be a primary physical meaning. The mapping from the rapidity $\psi$ onto the velocity $v$ is
given via hyperbolic tangent:
$$
v=c\tanh(\psi).
$$
The rapidity till up to now did not obtain its physical interpretation in the Lorentz-kinematics \cite{Leblond}.
However in the relativistic dynamics the hyperbolic angle $\psi$ can be expressed via external potential field
\cite{Yamal1}. It is relevant to point out on geometrical interpretation of the rapidity as a length in the hyperbolic
space of velocities. 

In the present paper we introduce a counterpart of the rapidity and explore its physical sense. Expressions for the
energy, momentum and the velocity via counterpart of the rapidity are regular at the point $m=0,~v=c$. The formulae for
the energy-momentum are represented as functions of some hyperbolic angle dual to the rapidity, by that reason we
suggest to denominate it as {\it counter-rapidity}. We shall see that the counter-rapidity has a certain physical
interpretation as a value inverse to the energy of the massless-state. The counter-rapidity has also its natural geometrical
interpretation as a distance in the space with hyperbolic metric.

In spite of formal reciprocity between rapidity and its counterpart, we shall see, they are quite
different. The hyperbolic angle $\chi$ dual to rapidity $\psi$ is equal to zero at $v=c$ and it tends to infinity
when $v$ goes to zero. It is shown that the hyperbolic angle $\chi$ can be presented as a fraction $\chi=m/\pi_0$,
where the variable $\pi_0$ in the limit $m=0$ becomes equal to the energy of a massless particle. This value can be
considered as a {\it counterpart} of the inertial mass. The dynamic situation of a moving particle is viewed as
follows: the particle with $m> 0$ is not able to attain the state of speed of light, conversely, the particle
possessing with $\pi_0> 0$ cannot fall to the rest state. In the rest state the energy is equal to the proper inertial
mass (in energy units) and in the opposite side, in the light-speed state, the energy is equal to the counterpart
of the mass, $\pi_0$. Thus, the kinetic energy of the relativistic particle is governed, beside the inertial mass $m$,
with its counterpart $\pi_0$. The massive particle moving with velocity less than light velocity possesses with both
types of mass.

In a similar manner as translations of the rapidity form a part of the Lorentz-group of transformations, translations of
the counter-rapidity also form some group of transformations. Geometrically, the concept of counter-rapidity directly
leads to geometrical interpretation of the relativistic dynamics within the framework of Poincar\'{e} model of
hyperbolic geometry where the counter-rapidity coincides with the length of the semi-circle in hyperbolic metrics.

The present theory does not require a new hypothesis. Nevertheless, this approach opens some new insights  and new ways
of generalization of the relativistic theory. One of the ways of generalization is open due to observation that the
formula for energy-momentum can be realized as a mapping from the energy of the massless state onto the massive one.
This formula looks like the well-known formula of "q-deformation". An analysis of this analogue prompts to introduce a
hypothesis on quantization of the velocity near the speed of light.

Usefulness of the present theory we demonstrate exploring the Dirac equation at the limit $m=0$. The representation for
the energy-momentum via counter-rapidity allows to reach a correct limit at $m=0$ explicitly displaying  violation of
the parity in the Dirac equation at this limit.

The paper is presented by the following sections.

In Section 2 concepts of the counter-rapidity and the counter-mass are introduced. The representation for the
energy-momentum via the counter-rapidity is deduced. In Section 3, the Dirac equation near $m=0$ is explored. In
Section 4, the formula for energy-momentum of massive particle is interpreted as a "q-deformation" of the state of a
massless particle. In Section 5 we build the group of transformations using the rapidity as a principal parameter of
transformation. In Section 6, the counter-rapidity is interpreted as a hyperbolic length of the semicircle of
Poincar\'e model of hyperbolic half-plane.

\section{ Representations of energy-momentum as functions of rapidity and its counterpart}

{\bf 2.1 Elements of relativistic dynamics of charged particle }

First of all let us remind some elements of the relativistic dynamics of charged particle necessary in subsequent
sections.

Consider a motion of the relativistic particle with charge $e$ in the external electromagnetic fields $\Ef$ and $\Bf$. The relativistic
equations of motion with respect to the proper time $\tau$ are given by the Lorentz-force equations \cite{Barut}:
$$
\frac{d\pf}{d\tau}=\frac{e}{mc}\Ef~{p_0}+\frac{e}{m}[\pf\times\Bf]),~~ \frac{dp_0}{d\tau}=\frac{e}{mc}(\Ef\cdot\pf), \eqno(2.1)
$$
$$
 \frac{d\rf}{d\tau}=\frac{\pf}{m},~~\frac{dt}{d\tau}=\frac{p_0}{mc}. \eqno(2.2)
$$
These equations imply the first integral of motion
$$
p_0^2-(\pf\cdot\pf)^2=M^2c^2. \eqno(2.3)
$$
In the case of stationary potential field, i.e. when $e\Ef=-{\nabla}V(r),$ the equations imply the other constant of motion, the energy of the
relativistic particle
$$
\ce=cp_0+V(r). \eqno(2.4)
$$
If the external electromagnetic field strengths is given in covariant form $F_{\mu\nu},~\mu,\nu=0,1,2,3$, then the Lorentz-force equation is written in the form
of Minkowski force-equation:
$$
\frac{d}{d\tau}u^{\mu}=\frac{e}{mc}{F^{\mu}}_{\nu}u^{\nu},~~u^{\mu}=\frac{dx^{\mu}}{d\tau}.\eqno(2.5)
$$
Correspondence with non-relativistic equations gives an interpretation of the constant of motion $M^2$ as a squared
mass of the particle, so that $M^2=m^2$. For the massless particle $M^2=0$. It is important to underline that the
relativistic dynamics of charged particle is formed by the pair of energies \cite{Yamal3}:
$$
q_1:=cp_0-mc^2,~~q_2:=cp_0+mc^2. \eqno(2.6)
$$
In the non-relativistic limit the former is transformed into kinetic energy of the Newtonian particle
$$ cp_0-Mc^2=cp_0-mc^2\rightarrow \frac{p^2}{2m}.\eqno(2.7)
$$

Next, we shall restrict ourselves by considering only lengths of the momenta. For that purpose let us project the Lorentz-force equations (2.1)
on the direction of motion
$$
\frac{dp}{d\psi}=~{p_0},~~ \frac{dp_0}{d\psi}= p,~~\frac{d\psi}{d\tau}=\frac{e}{mc}E,~ E=(\nf\cdot\Ef),~\nf=\frac{\pf}{p}.\eqno(2.8)
$$
From these equations it follows some dynamical interpretation of the rapidity $\psi$: the rapidity is integral
$\psi=e\int Ed\tau/mc$ of the force with respect to proper time of the particle measured along the trajectory. This
definition is in accordance with the definition given in Ref.\cite{Leblond}. From first two equations of (2.8) we find
$$
p_0=A(~\cosh(\psi)+B\sinh(\psi)~),~~p=A(~\sinh(\psi)+B\cosh(\psi)~). \eqno(2.9)
$$
Constants $A,B$ depend of the initial conditions. In addition, by substituting (2.9) into (2.3) we get
$$
A^2(1-B^2)=M^2c^2,
$$
and hence that  $B^2=1$ for massless particles. This condition is nothing else than the equation for the generator of a
hyperbolic complex number $z$ \cite{Fjelstad},\cite{Yaglom}, so that, if
$$
z:=x+{\e} y,~~{\e}^2=1,~~|z|=A,
$$
then,
$$
p_0=z,~p={\e}~z.
$$
In this context a question arises about representation of the momenta of a massive particles. The
answer to this question the reader may find in Refs.\cite{Yamal3}-\cite{Yamal4}.

Let $\psi=\psi_0$ for the rest state where $p=0,~p_0=mc$. Then Eqs.(2.9)are written as follows
$$
p_0=mc~\cosh(\psi-\psi_0),~~p=mc~\sinh(\psi-\psi_0). \eqno(2.10)
$$
Velocity with respect to coordinate time is defined by
$$
\frac{v}{c}=\frac{p}{p_0}=\tanh(\psi-\psi_0). \eqno(2.11)
$$
The same expression is used for mapping between rapidity $\psi$ and the velocity $v$ in the Lorentz-kinematics.

{\bf 2.2 Hyperbolic angle dual to rapidity}.

As usually formulae (2.10), (2.11) are considered as formulae of parametrization of the mass-shell equation (2.3). In
this context notice, however, that this is not unique form of parametrization. In fact, we can satisfy (2.3) by taking
$$
p_0=mc\coth{\chi},~~p=\frac{mc}{\sinh{\chi}}.\eqno(2.12)
$$
The parametrization as an objective does not give and interpretation of the parameter, however. Now, let us consider the
procedure of parametrization from another point of view.

Beforehand, let us remember one simple but useful formula \cite{Yamal2}:
$$
\frac{x_2}{x_1}=\frac{x_++x_-}{x_+-x_-}=\exp(x_-s),~~x_{\pm}=(x_2\pm x_1).       \eqno(2.13)
$$
An essential feature of this formula is that $x_-$ does not depend of translations of hyperbolic angle $s$.

Consider the quantities $q_2,q_1$ as solutions of quadratic equation
$$
X^2-2p_0X+p^2=0,  \eqno(2.14)
$$
where
$$
2p_0=q_1+q_2,~p^2=q_1q_2,~~2m=q_2-q_1.\eqno(2.15)
$$
According to Hamilton-Cayley theorem to equation (2.14) obeys the
following matrix
$$
E:= \left(
\begin{array}{cc}
0&-p^2\\
1&2p_0
\end{array} \right),
$$
eigenvalues of which are the roots of polynomial equation (2.14). Write the Euler formula for exponential function of
the matrix $E\phi$
$$
\exp(E\phi)=g_0(\phi;p_0,p^2) +E~g_1(\phi;p_0,p^2). \eqno(2.16)
$$
In terms of the eigenvalues this equation is decoupled  into pair of equations
$$
\exp(q_1\phi)=g_0(\phi;p_0,p^2)+x_1~g_1(\phi;p_0,p^2),~\exp(q_2\phi)=g_0(\phi;p_0,p^2)+x_2~g_1(\phi;p_0,p^2).
\eqno(2.17)
$$
Consider the following fraction
$$
\exp(2m\phi)= \frac{\exp(q_2\phi)}{\exp(q_1\phi)}=\frac{g_1q_2+g_0}{g_1q_1+g_0}= \frac{q_2-U}{q_1-U},
$$
where
$$
U=-\frac{g_0}{g_1}).\eqno(2.18)
$$
Thus, we get some interrelation between the fraction and the hyperbolic argument:
$$
\exp(2m\phi)=\frac{q_2-U}{q_1-U}.
$$
Now, we may fulfil translational transformation of the eigenvalues by $q_i=q_i+U,i=1,2$ remaining unchanged the value
of mass $2m=q_2-q_1$. This transformation induces a corresponding translation of the hyperbolic argument:
$$
\exp(2m(\phi+\delta(u)))= \frac{q_2}{q_1}.\eqno(2.19)
$$
Inversely, if we add to $\phi$ some value by $\phi=\phi+\delta$, this translation remains $m$ unchanged, whereas $p_0$ will undergo some translation $p_0=(q_2+q_1)/2-u$.
In this way we reach the conclusion that $m$ does not depend of $\delta$. Taking into account (2.15), write (2.19) as follows
$$
\frac{P_0+mc}{P_0-mc}=\exp(2m\phi),\eqno(2.20)
$$
where the mass $m$ is a constant of the evolution with respect to parameter $\phi$ whereas $P_0,P$ depend of $\phi$ and
this dependence is given by hyperbolic trigonometry
$$
P_0=mc\coth(mc\phi),~~P=\frac{mc}{\sinh(mc\phi)}.  \eqno(2.21)
$$
It is interesting to explore the use of the same formula  for the following fraction
$$
\frac{P_0+P}{P_0-P},
$$
where $P$ and $P_0$ are variables of the evolution. Let $P\neq 0$, then transform the fraction as follows
$$
\frac{P_0+P}{P_0-P}=\frac{\frac{P_0}{P}+1}{\frac{P_0}{P}-1}.
$$
Now we can use formula (2.13) to obtain the following equality
$$
\frac{\frac{P_0}{P}+1}{\frac{P_0}{P}-1}=\exp(2\psi). \eqno(2.22)
$$
Hence,
$$
\frac{P_0}{P}=\coth(\psi)=\frac{c}{v},~~\mbox{or},~~\frac{v}{c}=\tanh(\psi),\eqno(2.23)
$$
where $\psi$ is the rapidity.

We possess now with two different representations for the energy and momentum.
 The both given by hyperbolic trigonometry, the former is defined by formulae (2.10), (2.22),
 the latter is given by (2.21) where the hyperbolic angular is proportional to the mass of the particle.
An essential feature of the latter representation is its regularity
at the point $m=0,v=c$. At this point the hyperbolic argument is
equal to the energy-momentum of a massless particle:
$$
P(m=0)=P_0(m=0)=\frac{1}{\phi}=\pi_0.\eqno(2.24)
$$
Thus the value $c\pi_0$ is the energy of the relativistic system at the point $m=0,~~v=c$.
We should underline some differences between two hyperbolic angles.
At the rest, $\psi=0$, but $\phi=\infty$, and vice versa, when $v=c$, $\phi=0$ but
$\psi=\infty$. The particle with $m> 0$ is not able to attain the state of speed of light, conversely, the particle
possessing with $\pi_0> 0$ cannot fall to the rest state. In the rest state the energy is equal to the proper inertial
mass (in energy units) and, in the same manner, in the state of the light-speed the energy is equal to
$\pi_0$. Thus, the kinetic energy of the relativistic particle is governed, beside the inertial mass $m$,
with its counterpart $\pi_0$. The massive particle moving with velocity less than light velocity possesses with both
types of mass. The parameter $\pi_0$ is, in some sense, counterpart of the inertial mass which
determines the value of the kinetic energy of the motion. This quantity corresponds to the energy of the
particle in its massless state.

Let $v$ be the velocity of a particle with respect to coordinate time. This velocity is essentially less than the light
velocity, $v<c$. Beside $v$ let us introduce some complementary velocity $\bar{v}$ obeying the equation
$$
v^2+\bar{v}^2=c^2. \eqno(2.25)
$$
Now let us express $\bar{v}$ via the parameter ($\chi=mc\phi$). We get
$$
\bar{v}^2=c^2-v^2=c^2(1-\frac{P^2}{P_0^2})=c^2\tanh^2(\chi).\eqno(2.26)
$$
Substitute (2.23) and (2.26) into (2.25), this gives
$$
c^2\tanh^2(\psi)+c^2\tanh^2(\xi)=c^2. \eqno(2.27)
$$
Notice that $v_0$ is expressed via hyperbolic angle ($~\chi~$) by the same formula as $v$ is expressed via rapidity
$(\psi)$. Interrelation between $\chi$ and $\psi$ can be also expressed by the following formulae
$$
\exp(\psi)=\coth(\frac{\xi}{2}),~~\exp(\xi)=\coth(\frac{\psi}{2}).\eqno(2.28)
$$
These formulae are formulae of reciprocity. Hence, complementary velocity $v_0$ and counter-rapidity $\chi$ are reciprocal with
velocity $v$ and rapidity $\psi$.

\section{Dirac equation near zero-mass point}

The right- and left- two-component spinors under Lorentz-boost transformations are transformed as follows \cite{Waerden}
$$
\xi_{R}(P)=\frac{P_0+mc+(\sigma\cdot\vec P)~}{\sqrt{2m(P_0+mc)}}\xi_{R}(0),~~\xi_{L}(P)=\frac{P_0+mc-(\sigma\cdot\vec
P)~}{\sqrt{2m(P_0+mc)}}\xi_{L}(0),\eqno(3.1)
$$
where $\xi_{R}(0),~\xi_{L}(0)$ mean "right" and "left" spinors, correspondingly, at the rest state. When a particle stays at the rest it is
impossible to define its spin is "right", or is "left". Hence, $\xi_R(0)=\xi_L(0)$. From (5.1) it follows
$$
mc\xi_{R}(P)=(P_0+(\sigma\cdot\vec P)~)\xi_{L}(P),~~mc\xi_{L}(P)=(P_0-(\sigma\cdot\vec P)~)\xi_{R}(P).\eqno(3.2)
$$
These two equations can be written in a matrix form
$$
\left( \begin{array}{cc}
-mc&P_0+(\sigma\cdot\vec P)\\
P_0-(\sigma\cdot\vec P)&-mc
\end{array} \right)
\left( \begin{array}{c}
\xi_{R}\\
\xi_{L}
\end{array} \right)=0.\eqno(3.3)
$$
The differential form of these equations are given by
$$
\left( \begin{array}{cc}
-mc^2&i\hbar\partial_t-i\hbar c\vec\sigma\cdot\Delta\\
i\hbar\partial_t+i\hbar c\vec\sigma\cdot\Delta&-mc^2
\end{array} \right)
\left( \begin{array}{c}
\xi_R\\
\xi_L
\end{array} \right)=
\left( \begin{array}{c}
0\\
0
\end{array} \right).\eqno(3.4)
$$
This is Dirac equation for the massive particle with spin one-half in {\it chiral (or spinor, or Weyl)} representation \cite{Socolovsky}. This is
obtained from the standard representation using the unitary matrix
$$
S=S^{\dag}=S^{-1}=\frac{1}{\sqrt{2}}\left( \begin{array}{cc}
1&1\\
1&-1
\end{array} \right)\in U(4). \eqno(3.5)
$$
Two component spinors $\xi_R$ and $\xi_L$, respectively, correspond to the irreducible representations $(1/2,0)$ and
$(0,1/2)$ of the Lie algebra $SU(2) \bigoplus SU(2)$, which is isomorphic to the Lie algebra of the proper Lorentz
group. Under parity $(\vec x\rightarrow-\vec x)~$, $\xi_R$ and $\xi_L$ transform into each other. So the four-component
spinor $\Psi(chiral)$ is an irreducible representation of the Lorentz algebra extended by parity. Also $\left(
\begin{array}{c}
\xi_R\\
0
\end{array} \right) $ and
$\left( \begin{array}{c}
0\\
\xi_L
\end{array} \right) $ are eigenstates of the matrix $\gamma_5:=i\gamma_0\gamma_1\gamma_2\gamma_3$ with eigenvalues $+1$ and $-1$, repectively.
In the Weyl representation,
$\gamma_5(chiral)=\left( \begin{array}{cc}
1&0\\
0&-1
\end{array} \right)$.

For massless particles, the equations for the two-component spinors $\xi_R$ and $\xi_L$ decouple, and plane wave
solutions satisfy
$$
\widehat{\lambda}\xi_R=\xi_R,~~\widehat{\lambda}\xi_L=-\xi_L,
$$
where $\widehat{\lambda}$ is the helicity operator. These equations are called as the Weyl equations and historically
served as principal wave equations for "right" and "left" neutrino, respectively.  Both equations correspond to the
massless particles and the system of these equations are inter-related with operation of parity. Thus, violation of
symmetry of parity does not follow from the original equations, violation of parity in this approach is an additional
hypothesis. Now, let us explore the limit $m=0$ of the Dirac equation on making use of the representation of
energy-momentum via counter-rapidity.

Let us start from formulae (2.20-2.21) for the components of momentum as functions of the parameter $\phi$:
$$
\exp(2mc\phi)=\frac{P_0+mc}{P_0-mc},~P_0=mc~\coth(mc\phi),~~P=\frac{mc}{\sinh(mc\phi)},
$$
which can be re-written as follows
$$
(P_0+mc)\exp(-mc\phi/2)=P~\exp(mc\phi/2),~~(P_0-mc)\exp(mc\phi/2)=P~\exp(-mc\phi/2).\eqno(3.6)
$$
This system can be cast into  matrical form
$$
\left( \begin{array}{cc}
P_0+mc&P\\
P&P_0-mc
\end{array} \right)
\left( \begin{array}{c}
\exp(-mc\phi/2)\\
\exp(mc\phi/2)
\end{array} \right)=0.\eqno(3.7)
$$
Another equivalent form of these equations is given by the following system
$$
(P_0+P)=mc~\coth(\frac{mc\phi}{2}),~~(P_0-P)=mc~\tanh(\frac{mc\phi}{2}).
$$
These equations are written in one dimensional form where $P$ means a projection on the direction of motion:
$$
P=(\vec n\cdot\vec P),~~(\vec n\cdot\vec P)^2=P^2.
$$
Within the formalism of quantum mechanics of particle with one-half spin we write
$$
P\xi(P)=(\vec\sigma\cdot\vec P)\xi(P),~~(\vec\sigma\cdot\vec P)^2\xi(P)=P^2\xi(P),
$$
where $\vec \sigma=(\sigma_x,\sigma_y,\sigma_z)$ are the Pauli-sigma matrices, $\xi(P)$- two component Pauli- spinors.

Now, re-formulate one dimensional equations (3.3) in presence of the spin within the quantum framework \cite{Adan}. We
get
$$
(P_0+(\sigma\cdot\vec P))\xi_{L}(m,\phi)=mc\coth(mc{\phi})\xi_{L}(m,\phi),\eqno(3.7a)
$$
$$
(P_0-(\sigma\cdot\vec P))\xi_{R}(m,\phi)=mc\tanh(mc{\phi})\xi_{R}(m,\phi),\eqno(3.7b)
$$

Thus, the use of the counter-rapidity admits to split the Dirac equation into two equations for  two-component spinors.
In 1949 some interest came up in two-component wave equations (see, for instance, Refs.\cite{Jehle}) where had been
concluded that those wave equations in contrast to the four-component Dirac equations can only made covariant with
respect to the proper Lorentz group, not including covariance with respect to reflections.
In contrast to this case equatios (3.7a,b) are covariant with respect to reflections.
Beforehand notice, formulae (2.21) are invariant with respect to transformations
$$
\chi'=\chi+i2\pi,\eqno(3.8)
$$
which reflects the property of periodicity of the trigonometric finctions $\coth$ and $\sinh$. Under this transformation the functions
$\coth(\chi/2)$, $\tanh(\chi/2$ transform into each other providing symmetry of parity of Eqs.(3.7a,b).

The form of the Dirac equation presented by Eqs.(3.7a,b) allows us to reach correct limit in $m=0$. It should be
emphasized, in Dirac equations (3.3), (3.4) one cannot use $m=0$ directly because the Dirac equations are written for the particles
which admit a rest position. In particularly, this possibility was used in order to establish the equation
$\xi_R(0)=\xi_L(0)$ \cite{Ryder}. Moreover, in Eqs.(3.5) one cannot put $m=0$, too. In contrast to this case, equations
(3.7a,b) admit regular behavior at the limit $m=0$. In these equations by tending the mass $m$  to zero,  we arrive to
the following system of independent equations
$$
(P_0+(\sigma\cdot\vec P)~)\xi_L=2\pi_0~\xi_L,~~\mbox{and}~~(P_0-(\sigma\cdot\vec P))\xi_R=0.\eqno(3.9)
$$
These equations  are not more symmetric with respect to transformation of parity. The counterpart of the inertial mass $\pi_0$ allows
to realize the mass of the left-neutrino which experimentally is observed \cite{Marcino}. Notice, the mass of the
right-neutrino is equal to zero. The equations for the "left"-neutrino and the "right"- neutrino are not symmetric with
respect to parity.

For elementary particles, i.e. for the particles with quite small mass we can use $\phi=1/\pi_0$. Then, on making use
of $\phi=1/\pi_0$ in Eqs.(3.7a,b) we suggest to complete the Dirac equation for the electron as follows
$$
(P_0+(\vec P\cdot\sigma)~)\Psi_1=mc~\coth \frac{mc}{{\pi_0}}\Psi_1,~~(P_0-(\vec P\cdot\sigma)~)\Psi_2=mc~\tanh
\frac{mc}{{\pi_0}}\Psi_2,\eqno(3.10)
$$
where $\pi_0$ is a mass of the neutrino.

\section{Mapping from the massless state onto the massive one as a q-deformation }

Near the zero-mass point $P_0 \simeq \pi_0$, hence, we may approximate the energy as follows: $P_0=K\pi_0$. In this way
we come to following equation
$$
\frac{mc}{K\pi_0}=\tanh\frac{mc}{\pi_0}. \eqno(4.1)
$$
There are two cases to consider.

Suppose first that $K < 1$. Then, the equation (4.1) has only one solution, namely $m=0$.

Suppose now that $K > 1$. Since for small values of the mass we can use the expansion
$$
\tanh y=y-\frac{1}{3}y^3,
$$
we come to cubic equation
$$
\frac{1}{K}y=y-\frac{1}{3}y^3,~~y=\frac{mc}{\pi_0},
$$
with solutions
$$
y=\pm\sqrt{3}\sqrt{1-\frac{1}{K}}~\mbox{and}~y=0.
$$
Consequently, for the quite small values of the mass, the equation (4.1) gives three possible quantities for the
masses:
$$
(1)~m=0,~~(2)~ m=\pm \frac{\pi_0}{c}\sqrt{3}\sqrt{1-\frac{1}{K}}.\eqno(4.2)
$$
The first one corresponds to the massless particle. However, it is seen that this is not unique solution of equation
(4.1) near the zero-mass point. The equation admits the other solution given by finite value. We suggest a hypothesis
to consider this quantity as a mass of an elementary particle.

Near the zero-mass point we may take $\phi={1}/{\pi_0},$ then, formulae for the energy-momentum (2.21) are written as follows
$$
P=\frac{mc}{\sinh(\frac{mc}{\pi_0})},~~P_0=mc\coth(\frac{mc}{\pi_0}).\eqno(4.3)
$$
Notice, these formulae  can be considered as a mapping for the energy-momenta from the massless state onto the state with finite mass. For the
massless particle, the photon, the following well-known relationship between length of the wave and the momentum holds true
$$
\lambda=\frac{h}{\pi_0},~~\pi_0=\frac{h}{\lambda}.
$$
By using these formulae in (4.4) we come to the following mapping
$$
cP_0=mc^2~\coth(\frac{mc^2}{h\nu}),~~P=\frac{mc}{\sinh(\frac{mc}{h}\lambda)}.
$$
When $m\rightarrow 0$  these formulae reduce to covariant de Brogli\`{e} formulae
$$
cP_0(m_e)=h\nu,~~P(m_e)=\frac{h}{\lambda}.\eqno(4.4)
$$
where $m_e$ is the mass of the electron. We must keep in mind, however that the de Brogli\`{e}'s formulae are valid
only for the masses of elementary particles, i.e. when the mass of the particle is quite small.

Now let us explore some modification of formulae (2.21). For that purpose introduce some parameter in unit of mass and
label this parameter by $\kappa$. Define a dimensionless variable $\alpha$ by
$$
\alpha=\phi{\kappa c},~~ \phi=\frac{\kappa c}{\alpha}.
$$
$\alpha$ runs from $\alpha=0$ (which corresponds to velocity $v=c$) till $\phi=\infty$ ( which corresponds to the rest
state with $v=0$).

Re-write formulae for the energy-momentum (2.21) in these variables
$$
P=\frac{mc}{\sinh(\frac{m}{\kappa}\alpha)},~~ P_0={mc}~{\coth(\frac{m}{\kappa}\alpha)}. \eqno(4.5)
$$
For velocity we get
$$
\frac{v}{c}=\frac{1}{\cosh(\frac{m}{\kappa}\alpha)}.\eqno(4.6)
$$
This formula can be considered as some mapping between dimensionless parameters $v/c$ and $\alpha$.

Let us examine physical sense of the new parameter $\kappa$.

Here it is useful to notice that the formula for the momentum admits the following integral representation
$$
\frac{\kappa}{P}=\frac{\sinh(\frac{m}{\kappa}\alpha)}{\frac{m}{\kappa}}=\int^{\alpha/2}_{-\alpha/2}~\exp(2\frac{m}{\kappa}x)dx.
\eqno(4.7)
$$
 In \cite{Yamal1} it has been shown that $\kappa^{-1}$ geometrically can be interpreted as a curvature of a hyperbolic space. In this space the length
 of the circle with radius $m$ is defined by formulae \cite{Kagan}:
$$
L(\alpha=1):=2\pi\kappa\sinh(\frac{m}{\kappa}).\eqno(4.8,a)
$$
Correspondingly, the length of the circle with radius  $m\alpha$ is equal
$$
L(\alpha)=2\pi\kappa\sinh(\frac{m}{\kappa}\alpha).\eqno(4.8,b)
$$
Taking into account this correspondence let us perform the following modifications in the formulae for energy-momentum
$$
\frac{mc}{P}=\sinh(\frac{m}{\kappa}\alpha)\rightarrow
\frac{c\kappa}{P}=\frac{\sinh(\frac{m}{\kappa}\alpha)}{\sinh(\frac{m}{\kappa})}. \eqno(4.9)
$$
$$
\frac{P_0}{mc}=\coth(\frac{m}{\kappa}\alpha) \rightarrow
\frac{P_0}{c\kappa}=\sinh(\frac{m}{\kappa})\coth(\frac{m}{\kappa}\alpha). \eqno(4.10)
$$
 Now remember the formula of q-deformation of an integer quantity $N$:
$$
(N)_q:=\frac{q^{N}-q^{-N}}{{q}-q^{-1}}.\eqno(4.11)
$$
From this point of view the fraction in (4.9) is q-deformation of $\alpha$ with parameter of deformation
$q=\exp(m/\kappa)$. In notations of (4.11) Eq.(4.9) can be written as follows
$$
\frac{c\kappa}{P}=(\alpha)_q,~~q=\exp(\frac{m}{\kappa}).\eqno(4.12)
$$
Notice,  $(\alpha=1)_q=1$  for any $q$. There fore
$$
P(m\neq 0,\alpha=1)=\kappa c,~~\mbox{for~any}~m>0.
$$
On the other hand, if $m=0$ and $q=1$ then from (4.12) it follows
$$
P(m=0)=\pi_0=\frac{c\kappa}{\alpha}. \eqno(4.13)
$$
Hence at the point $\alpha=1$ momenta of the particles with different masses, including the massless particle, are equal to $c\kappa$:
$$
P(m\neq 0,\alpha=1)=P(m=0,\alpha=1)=\pi_0(\alpha=1)=c\kappa.\eqno(4.14)
$$
In fact, these formulae imply existence of a point on the axis of momentum where the momenta of the massive and massless particles are equal.
Notice, however, the velocity of the particle at this point is not equal to the light velocity. At this point the energy and the velocity are
given by
$$
cP_0(\alpha=1)=c^2\kappa\cosh~(\frac{m}{\kappa}),~~~ v=\frac{c}{\cosh~(\frac{m}{\kappa})}.\eqno(4.15)
$$
Thus, the point $\alpha=1$ is a peculiar point of the relativistic dynamics where the constant $\kappa$ is now has
to be understood as an universal constant.

The procedure of $q$-deformation is usually used in order to extend formulae obtained for integer number to the field of real numbers.
Seemingly, $\alpha$ is presented by integer numbers. Here let us give an additional argumentation to this hypothesis. For that purpose remember on
integral representation (4.7). Now, instead of the fraction in (4.7) we deal with the fraction
$$
\frac{c\kappa}{P}=\frac{\sinh(\frac{m}{\kappa}\alpha)}{\sinh(\frac{m}{\kappa})}\eqno(4.16)
$$
It is interesting to observe, for the fraction in (4.16) we shall put an equivalence a sum instead of the integral because now we assumed that $\alpha$ is an
integral number. Let $J$ be a half-integer number with $J=0,1/2,1,3/2,2,...$, and $\alpha=2J+1=1,2,3,...$. Then the following equation holds
true
$$
\frac{c\kappa}{P}=\frac{\sinh(\frac{m}{\kappa}(2J+1))}{\sinh(\frac{m}{\kappa})}=\sum^J_{n=-J}\exp(n\frac{m}{\kappa}).
$$
This equality prompts us to introduce a hypothesis on quantization of $\alpha$. Experimentally the quantization can be observed near the light
velocity where the velocity of the massive particle brings nearer the light velocity spasmodically according to law
$$
v=\frac{c}{\cosh(\frac{m}{\kappa}(2J+1))}.
$$

By comparing (4.13) with  de Broglie formulae (4.4), we get
$$
\lambda=\frac{h\alpha}{\kappa c}.
$$
Substituting here the quantized value of $\alpha$ we obtain quantization of the wave-length:
$$
\lambda=\frac{h}{\kappa c}(2J+1).
$$

\section{ Covariant form of the counter-rapidity and its group of transformation }

In the Lorentz-group rapidity fulfils function of the parameter of transformation corresponding to the Lorentz- boosts.
In this form the rapidity is presented as a vector with three components.
In this section we shall construct the group which uses the counter-rapidity as a parameter of transformation. Let us start
with evolution equations for momenta with respect to rapidity
$$
\frac{d p}{d\phi}=- p~{p_0},~~ \frac{dp_0}{d\phi}=-p^2.\eqno(5.1)
$$
Remembering  that the momentum is three-dimensional Euclidean vector this equation can be extended as follows
$$
\frac{d\vec p}{d\phi}=-\vec p~{p_0},~~ \frac{dp_0}{d\phi}=-(\vec p\cdot \vec p).\eqno(5.2)
$$
However, we still did not reach Lorentz-covariant form of the evolution equation.

Now, let us consider transformation for two components of the four-velocity $u_{\mu}:~u_0,u$ with respect to translation of
hyperbolic angle $\chi$: $ \chi'=\chi+\delta$ . Introduce the following variables
$$
V_0={\cosh\delta},~~V=\frac{c}{\sinh\delta}.
$$
Then, by using hyperbolic trigonometry we get
$$
u_0'=c \coth(\chi+\delta)=\frac{u_0V_0+c^2}{u_0+V_0},~~u'=\frac{c}{\sinh(\chi+\delta)}=uV\frac{1}{u_0+V_0}. \eqno(5.3)
$$
Let us underline some features of the transformation. At the rest state we have $u_0=c,u=0$. According to (5.3), we get
$$
u_0'=\frac{c~V_0+c^2}{c+V_0}=c,~~u'=\frac{c}{\sinh(\chi+\delta)}=0~V\frac{1}{c+V_0}=0.
$$
These formulae mean that the rest state is a {\it fixed point} for transformations (5.3). For the complementary velocity we write
(see, (2.26)):
$$
\bar{v}'=\frac{\bar{v}+\bar{V}}{1+\frac{\bar{v}\bar{V}}{c^2}},~~\bar{V}=c\tanh(\delta).
$$
At the rest state, $\bar{v}=c$. In this case
$$
\bar{v}'=\frac{c+\bar{V}}{1+\frac{c\bar{V}}{c^2}}=c.
$$
At the light-speed state, $\bar{v}=0$, then
$$
\bar{v}'={0+\bar{V}}=\bar{V},~~
$$
what means that for this transformation the state of the light-speed is relative.

Notice, in order to employ formula (5.3) for a four-vector with coordinates $x_0,x$ one has to use the length  of
the vector. Define the square of the length by
$$
\rho^2=x_0^2-x^2.\eqno(5.4)
$$
Then formulae of transformation (5.3) for four-vector $x_0,x$ are written as follows
$$
x_0'=\frac{x_0r_0+\rho^2}{x_0+r_0},~~x'=\frac{x~r}{x_0+r_0}.\eqno(5.5)
$$
Here notice that the parameter of transformation is given via coordinates $r_0,r$. In general case, suppose that
the  hyperbolic angle $\phi$ is only one of the components of four-component
vector $\xi_{\nu},\nu=0,1,2,3$.  With respect to these parameters the evolution of four-vector $x_{\mu},\mu=0,1,2,3$ is
governed by the following equations
$$
\frac{\partial x_{\nu}}{\partial \xi_{\mu}}=\rho^2~\eta_{\nu\mu}-x_{\nu}x_{\mu},~~ \rho^2=(x^{\mu}x_{\mu}).\eqno(5.6)
$$
In view of equation (5.6), covariant form of evolution equations (5.2) for the four-vector of momentum are given by
$$
\frac{\partial P_{\nu}}{\partial \xi_{\mu}}=(mc)^2\eta_{\nu\mu}-P_{\nu}P_{\mu}.
$$
Now, let $\mu=0$. Then
$$
\frac{\partial P_{\nu}}{\partial \xi_{0}}=(mc)^2\eta_{\nu\mu}-P_{\nu}P_{0},
$$
which is decoupled into two equations
$$
(1)~~\frac{\partial P_{0}}{\partial \xi_{0}}=(mc)^2-P_{0}^2=-P^2;~~(2)~~
\frac{\partial \vec P}{\partial \xi_{0}}=-\vec P~P_{0}.
$$
We recover Eqs.(5.2) supposing that $\xi_0=\phi$.

The generators of transformation with respect to parameters $\xi_{\nu},\nu=0,1,2,3$ are defined by derivatives
$$
G_{\nu}=\frac{\partial }{\partial \xi_{\nu}},~~\nu=0,1,2,3.\eqno(5.7)
$$
Express these generators in terms of four-vector $x_{\mu}$ on making use of (5.6). This leads to
$$
G_{\nu}=\frac{\partial x_{\mu} }{\partial \xi_{\nu}}\frac{\partial }{\partial x_{\mu}}= \rho^2\frac{\partial }{\partial
x_{\nu}}-x_{\nu}~(~x_{\mu} \frac{\partial}{\partial x_{\mu}}),\nu,\mu=0,1,2,3.\eqno(5.8)
$$
Now let us construct commutation relations containing the generators $G_{\mu},\mu=0,1,2,3$ as elements of the
group which we are seeking. Let us denominate this group by "$\Gamma$-group".  First of all calculate commutators between $G$-generators. They are
$$
[G_{\mu},G_{\nu}]=\rho^2~M_{\mu\nu},~~M_{\mu\nu}=x_{\mu}\partial_{\nu}-x_{\nu}\partial_{\mu}.\eqno(5.9)
$$
Thus, in surplus $\Gamma$-group contains elements of the Lorentz-group $M_{\mu\nu}$ and the factor $\rho^2$ which is
also the element of the $\Gamma$-group. The length is invariant under action of $G$- generator and the
generators of the Lorentz-group,
$$
[G_{\mu},\rho^2]=0,~~[\rho^2,M_{\mu\nu}]=0.\eqno(5.10)
$$
The operator $G_{\mu}$ is a sum of two operators, namely, the differential operator in Minkowski space,
$\partial_{\mu}$ and generator of dilatation $D=(x^{\mu}\partial_{\mu})$:
$G_{\mu}=\rho^2\partial_{\mu}-x_{\mu}~D.$
By taking into account $ [M_{\mu\nu},D]=0$ and $ [\rho^2,M_{\mu\nu}]=0$, we get
$$
[M_{\mu\nu},G_{\lambda}]=\eta_{\nu\lambda}G_{\mu}-\eta_{\mu\lambda}G_{\nu}.\eqno(5.11)
$$
The generators of the Lorentz-group obey ordinary commutation relations
$$
[M_{\mu\nu},M_{\lambda\eta}]=(~\eta_{\mu\lambda}M_{\nu\eta}-\eta_{\nu\lambda}~M_{\mu\eta}+
\eta_{\mu\eta}M_{\lambda\nu}-\eta_{\nu\eta}~M_{\lambda\mu}~).\eqno(5.12)
$$
The $\Gamma$-group possesses with two Casimir operators:
$$
C_1=G^2,~~C_2=M_{\mu\nu}M_{\mu\nu}~G^2/2-M_{\mu\lambda}M_{\mu\lambda}G_{\mu}G_{\nu}.\eqno(5.13)
$$
The $\Gamma$- group  can be extended by introducing the operator dilation $D=(x^{\mu}\partial_{\mu}).$ In this
case we must add two additional commutators
$$
[D,G_{\mu}]=G_{\mu},~~[D,\rho^2]=2\rho^2.
$$
This extension similar the extension of  Poincar\'e group till the Weyl group by adding the dilation operator
\cite{Konopol}.

It is worth to note an analogue between elements of special conformal group $K_{\mu},\mu=0,1,2,3$ and the
generators of $\Gamma$-group. A representation of the special conformal elements via differential operators in Minkowski space is given by
$$
K_{mu}=\rho^2\partial_{\mu}-2x_{\mu}~D.
$$
This operator differs of the operator $G_{\mu}$ only with factor $2$ at the second term. Also, let us underline similarity of the generator $G_{\mu}$ with the generator of translation on the surface with constant
curvature \cite{Yamal5}. In fact, if the factor $\rho^2=constant$ the operator $G$ is transformed into the operator of
translation along hyperbolic surface imbedded into four-dimensional Minkowski space.

Furthermore, let us emphasize some analogue between generator $G$ and the generator of translation of the
Poincar\'{e}-group. The following formulae exhibit this analogue:
$$
[G_{\nu},x_{\mu}]=\rho^2~\eta_{\nu\mu}-x_{\nu}x_{\mu},~~ \frac{\partial x_{\mu}}{\partial
\xi_{\nu}}=\rho^2~\eta_{\mu\nu}-x_{\mu}x_{\nu},
$$
with
$$
[ p_{\nu},x_{\mu}]=~\eta_{\nu\mu}= \frac{\partial x_{\mu}}{\partial x_{\nu}}.
$$
In conclusion let us notice that the commutation relations of the $\Gamma$-group can be realized by the set of finite dimensional
matrices formed by Dirac {\it gamma}- matrices \cite{Keller}. These matrices satisfy anti-commutation relations
$$
\{ \gamma_{\mu},\gamma_{\nu}~\}=\gamma_{\mu}\gamma_{\nu} +\gamma_{\nu}\gamma_{\mu}=2~\eta_{\mu\nu}~1,
$$
where  $ \eta_{\mu\nu}=diag(1,-1,-1,-1)$ is the Minlowski metric. $\gamma_0$ is Hermitian: $\gamma_{0}^{\dag}=
\gamma_0$, while the $\gamma_{k},k=1,2,3$ are anti-Hermitian: $ \gamma_{k}^{\dag}=-\gamma_{k}$.

Within the algebra of $\gamma$-matrices we can construct finite dimensional realization of the commutation relations (5.9),(5.11),(5.12):
$$
[\gamma_{\mu},\gamma_{\nu}]=~\Sigma_{\mu\nu},~~~
[\Sigma_{\mu\nu},\gamma_{\lambda}]=\eta_{\nu\lambda}\gamma_{\mu}-\eta_{\mu\lambda}\gamma_{\nu},
$$
$$
[\Sigma_{\mu\nu},\Sigma_{\lambda\eta}]=(~\eta_{\mu\lambda}\Sigma_{\nu\eta}-\eta_{\nu\lambda}~\Sigma_{\mu\eta}+
\eta_{\mu\eta}\Sigma_{\lambda\nu}-\eta_{\nu\eta}~\Sigma_{\lambda\mu}~).
$$
With the spin-matrices $\Sigma_{\mu\nu}$ one may organize finite-dimensional non-unitary representation of the
Lorentz-group. In this context the set of matrices $ \gamma_{\mu},~\Sigma_{\mu\nu}$ will realize non-unitary
finite-dimensional representation of the $\Gamma$-group. Notice, only compact groups admit  finite non-trivial unitary
representations \cite{Konopol}.

\section{The counter-rapidity as a distance in the space with hyperbolic metric}

In Ref.\cite{Yamal3} it has been shown that the evolution generated by matrix $E$ (i.e. by mass-shell equation)
geometrically can be depicted as a rotational motion of straight line around semicircle to which the line is tangent
(see, Theorem 2.1 in p.1051 \cite{Yamal3}). Such line is described by by hyperbolic trigonometry. The hyperbolic angle $\chi=m\phi$ is measured from the top-point of the semicircle (see, Fig.1
in p.1051 \cite{Yamal3})(throughout this section we take $c=1$).  This semicircle is nothing else than a geodesic line in the Poincar\'e model of the
hyperbolic plane.

The model of the hyperbolic plane means a choice on an underlining space and a choice of how to represent basic geometric objects, such as
points and lines, in this underlining space. There are a large number of possible models for the hyperbolic plane, we shall work in a single
specific model. The model of the hyperbolic plane is the {\it upper half-plane} model. The underlining space of this model is the upper half-plane $H$ in the
complex plane $C$ defined to be
$$
H=\{ z\in C:Im(z)>0 \}.
$$
It is used the usual notion of point that $H$ inherits from $C$. It is also used the usual notion of angle that $H$ inherits from $C$, that is,
the angle between two curves in $H$ is defined to be the angle between the curves when they are considered to be curves in $C$, which in turn is
defined to be the angle between their tangent lines.

There two seemingly different types of {\it hyperbolic line}, both defined in terms of Euclidean objects in $C$. One is the intersection of $H$
with a Euclidean line in $C$ perpendicular to the real axis $R$ in $C$. The hyperbolic lines should have one property, namely, there should
always exist one and only one hyperbolic line between any pair of distinct points of $H$.

{\bf The definition of Cross Ratio}.

Given four distinct points $z_1,z_2,z_3,z_4$ in $C$, define the {\it cross ratio} of these points to be
$$
[z_1,z_2;z_3,z_4]=\frac{(z_1-z_4)}{(z_1-z_2)}\frac{(z_3-z_2)}{(z_3-z_4)}.\eqno(6.1)
$$

{\bf Theorem} (see, \cite{Katok},p.12).

Let $z,w \in H (z\neq w)$, and let $z^{*}$ and $w^{*}$ are end-points of a geodesic line passing through $z,w$ belonging $R\bigcup\{\infty\}$,
and chosen in a such way that $z$ lies between $z^{*}$ and $w$. Then
$$
\rho(z,w)=\ln[w,z^{*};z,w^{*}].\eqno(6.2)
$$
This function is invariant under M\"{o}b(+) transformation.

In Section 2 (formula (2.18)) we have shown that the evolution of the energy-momentum with respect to parameter $\phi$
is determined by function $U$:
$$
P_0(\phi)=P_0(\phi_0)-U(\phi,\phi_0),\eqno(6.3)
$$
where
$$
U=-\frac{g_0}{g_1},\eqno(6.4)
$$
where $g$-functions satisfy the equation  \cite{Yamal4}
$$
g_0(g_0+2p_0g_1)+p^2g_1^2=\exp(a_1\phi).
$$
Divide this equation by $g_1^2$,
$$
(\frac{g_0}{g_1})^2+2p_0\frac{g_0}{g_1}+p^2=\exp(2p_0\phi)\frac{1}{g_1^2},
$$
and by using (6.4), we get
$$
U^2-2p_0U+p^2=\exp(2p_0\phi)\frac{1}{g_1^2}.\eqno(6.5)
$$
Now, calculate derivative of $U$ on making use of formulae of derivation for $g$-functions \cite{Yamal4}:
$$
\frac{d}{d\phi}\left( \begin{array}{c}
g_0\\
g_1
\end{array} \right)=
\left( \begin{array}{cc}
0&-p^2\\
1&2p_0
\end{array} \right)
\left( \begin{array}{c}
g_0\\
g_1
\end{array} \right).
$$
In this way we prove that
$$
\frac{dU}{d\phi}=\exp(2p_0\phi)\frac{1}{g_1^2}.\eqno(6.6)
$$
Substituting (6.6) into (6.5) we come to the Ricatti equation for function $U$:
$$
U^2-2p_0U+p^2=\frac{dU}{d\phi}.\eqno(6.7)
$$
Like tangent (or, cotangent) function, the function $U=U(\phi)$ is found as an inverse of the function obtained as a
result of calculation of the integral
$$
\int^U_W~\frac{dx}{x^2-2p_0x+p^2}.\eqno(6.8)
$$
Let $x_1,x_2$ be roots of the polynomial $F(x):=x^2-2p_0x+p^2=(x-x_1)(x-x_2)$. Then, $2m=x_1-x_2,~~2p_0=x_1+x_2.$
Integral (6.8) is a table integral of the form
$$
\int^U_W~\frac{dx}{x^2-2p_0x+p^2}=
\frac{1}{2m} \ln\frac{U-x_1}{W-x_1}-\frac{1}{2m}\ln \frac{U-x_2}{W-x_2}=
=\frac{1}{2m} \ln\frac{U-x_1}{W-x_1}\ln \frac{W-x_2}{U-x_2}.\eqno(6.9)
$$
Now let us keep only the first part of the answer depending on the first limit of the integral:
$$
\phi:=\log[\frac{U-x_1}{U-x_2}]^{\frac{1}{2m}}.
$$
Inverting this logarithm function we come to the following exponential function
$$
\exp(2m\phi)=\frac{U-x_1}{U-x_2}.
$$
By taking into account (2.7), we may write this formula as follows
$$
\exp(2m\phi)=[\frac{x_1-U}{x_2-U}]=[\frac{p_0-U+m}{p_0-U-m}].
$$
From this formula we get:
$$
p_0-U(\phi)=\coth(m\phi),~~U(\phi)=p_0-m\coth(m\phi),
$$
Thus, we again recover formulae (2.20) (with $c=1$).

Compare the formula for Cross Ratio (6.1) with  the result of calculation of the integral (6.9). For the former we have,
$$
\rho(z,w)=\ln[w,z^{*};z,w^{*}]=\ln~\frac{(w-z^{*})}{(z-z^{*})}\frac{(z-w^{*})}{(w-w^{*})},\eqno(6.10)
$$
and for the latter,
$$
2m\int^w_z~\frac{dx}{x^2-2p_0x+p^2}=\ln\frac{w-z^{*}}{z-z^{*}}   \frac{z-w^{*}}{w-w^{*}},
$$
where $z^{*},w^{*}$ are roots of the polynomial $f(x):=x^2-2p_0x+p^2.$ Consequently,
$$
2m\int^w_z~\frac{dx}{x^2-2p_0x+p^2}=\rho(z,w).
$$
In this wa we come to the conclusion that the
hyperbolic  distance between two points  on the hyperbolic plane is defined with the integral (6.8).

\section{ Conclusions}

In order to complete our knowledge it is important to make observations of the physical objects from different points
of references. Up to now, the relativistic physics dealt only with observers installed upon inertial systems of
references with velocities less than light velocity. The particle with non-trivial inertial mass accelerating tends to the
speed of light, but never attains this velocity. For inertial observers this point is forbidden prohibited by convention.

In the present theory we suggest go from the light-speed state to the rest state and use the velocity complementary to the velocity
of the particle (observer). For the latter we put in correspondence some hyperbolic angle, we interpret as counterpart of the rapidity.
The formulae for the energy and momentum via counte-rapidity  are regular at the point $m=0,~v=c$. 
The counter-rapidity has a certain physical
interpretation as the value inverse to the energy of a massless-state. This value has also its natural geometrical
interpretation as a distance in the space with hyperbolic metric.

In spite of formal reciprocity between rapidity and its counterpart, they are quite
distinct. The hyperbolic angle $\chi$ dual to rapidity $\psi$ is equal to zero at $v=c$ and it tends to infinity
when $v$ goes to zero. It is shown that the hyperbolic angle $\chi$ is proportional to the proper mass $\chi=mc^2/\pi_0$,
where the variable $\pi_0$ in the limit $m=0$ becomes equal to the energy of a massless particle, we denominate $\pi_0$
as a counterpart of the inertial mass. The dynamic situation of a moving particle is viewed as
follows: the particle with $m> 0$ is not able to attain the state of speed of light, conversely, the particle
possessing with $\pi_0> 0$ cannot fall to the rest state. In the rest state the energy is equal to the proper inertial
mass (in energy units) and in the opposite side, in the light-speed state, the energy is equal to the counterpart
of the mass, $\pi_0$. Thus, the kinetic energy of the relativistic particle is governed, beside the inertial mass $m$,
with its counterpart, $\pi_0$. The massive particle moving with velocity less than light velocity possesses with both
types of mass: $m$ and $\pi_0$.

In a similar manner as translations of the rapidity form a part of the Lorentz-group of transformations, translations of
the counter-rapidity form some group of transformations. Geometrically, the concept of counter-rapidity directly
leads to geometrical interpretation of the relativistic dynamics within the framework of Poincar\'{e} model of
hyperbolic half-plane where the counter-rapidity coincides with the length of the semi-circle in hyperbolic metrics.

According to L\'evy-Leblond \cite{Leblond}, the rapidity till up to now did not obtain its
physical interpretation in the Lorentz-kinematics. The rapidity as an hyperbolic angle has been introduced in order to
use the hyperbolic trigonometry in the relativistic kinematics. In the sequel, the rapidity obtained its geometrical
interpretation as a length of the geodesic line in the hyperbolic geometry. The link between  hyperbolic geometry of
Lobachevsky-Gauss-Bolyai and the relativistic kinematics firstly had been discovered independently by V.Vari\v{c}ak
\cite{Varicak}, H.Herglotz \cite{Herglotz} and A.A.Robb \cite{Robb}. Later, this idea was developed by A.P.Kotelnikov
from the university in Kazan, where N.Lobachevski worked out his geometry \cite{Chernikov}. Contemporary model of the
link between relativistic kinematics and the hyperbolic geometry based on non-associative algebras and the concept of
gyrogroups has been constructed by A.Ungar \cite{Ungar}. In the scope of the quantum relativistic dynamics the
hyperbolic geometry intensively was exploited within so-called "quasi-potential models of the elementary particles"
(see, for instance, \cite{Kad}, \cite{Mirkas} and references therein).

\end{document}